\documentclass[]{raa}            
\usepackage{graphicx,times}
\usepackage{natbib}

\providecommand{\bm}[1]{\mbox{\boldmath$#1$\unboldmath}}

\begin{document}

   \title{The physical origin of optical flares following GRB 110205A and the nature of the outflow}

 \volnopage{ {\bf 20xx} Vol.\ {\bf xx} No. {\bf XX}, 000--000}
   \setcounter{page}{1}

   \author{Wei-Hong Gao
      \inst{1}  }

   \institute{Department of Physics and Institute of Theoretical
Physics, Nanjing Normal University, Nanjing, 210008, China; {\it
gaoweihong@njnu.edu.cn} }

\abstract{The optical emission of GRB 110205A is distinguished by
two flares. In this work we examine two possible scenarios for the
optical afterglow emission. In the first scenario, the first optical
flare is the reverse shock emission of the main outflow and the
second one is powered by the prolonged activity of central engine.
We however find out that it is rather hard to interpret the late
($t>0.1$ day) afterglow data reasonably unless the GRB efficiency is
very high ($\sim 0.95$). In the second scenario, the first optical
flare is the low energy prompt emission and the second one is the
reverse shock of the initial outflow. Within this scenario we can
interpret the late afterglow emission self-consistently. The reverse
shock region may be weakly magnetized and the decline of the second
optical flare may be dominated by the high latitude emission, for
which strong polarization evolution accompanying the quick decline
is possible, as suggested by Fan et al. in 2008. Time-resolved
polarimetry by RINGO2-like polarimeters will test our prediction
directly.  }

   \authorrunning{Wei-Hong Gao }
   \titlerunning{The optical flares of GRB 110205A }
      \maketitle

\section{Introduction}
\label{sect:intro}

GRB 110205A was triggered and located by the {\it Swift} Burst Alert
Telescope(BAT) at 02:02:41 UT and the BAT lightcurve showed activity
with multiple peaks ending at 1500 seconds after the GRB trigger,
with a peak count rate of 4500 counts/s (15-150 keV) at 210 seconds
after the trigger. The duration of GRB 110305A is $T_{90}=257 \pm
25 $ seconds in the range from 15 keV to 350 keV (Beardmore et al.
2011a, Markwardt et al. 2011).

The redshift of GRB 110205A was $z=2.2$ through detecting and
analyzing the optical spectrum of the host galaxy (Cenko et al.
2011). As observed by Konus-Wind, the isotropic energy is $E_{\rm
iso} = (4.34 \pm 0.42)\times10^{53}$ erg, using a standard cosmology
model with $H_0 = 71 {\rm km~s^{-1}~Mpc^{-1}}$, $\Omega_{\rm M}=
0.27$, and $\Omega_{\Lambda} = 0.73$ (Gorosabel et al. 2011).

The optical observation started at 90.6 seconds after the GRB110205A
trigger (Klotz et al. 2011a). Two optical flares were detected during
the observation. The optical brightness increased to the first peak
of ${\rm R}\sim 16.7$ about 225 seconds after the trigger. Then the flux
decreased at ${\rm R}\sim 18$ about 360 seconds after the trigger (Klotz et
al. 2011b). The brightness increased again, and the second peak was
${\rm R}\sim 13.7$ at about 1140 seconds after the trigger(Andreev et al.
2011). The lightcurve of the optical flares in R-band is illuminated
in Fig.1. By fitting to this lightcurve a power law $F_{t}\propto
t^{-\alpha}$, we find $\alpha_{\rm I} \sim 2.5$ for the decline of the
first flare. And $\alpha_{\rm IIr}\sim -4$ for the rising phase of the
second flare. For the second flare decline we can obtain
$\alpha_{\rm IId}\sim2.4$ until break at $t\sim8000$ seconds, then it
decayed as $F_{t}\propto t^{-1.6}$.

The XRT began observing GRB 110205A at 02:05:16.8 UT, 155.4 seconds
after the BAT trigger(Beardmore et al. 2011a). The X-ray lightcurve
comprises a number of flares followed by a power-law decay. The
late-time lightcurve can be modelled with  a power-law decay with an
index of $\alpha_{\rm x}=1.63\pm0.10$ from the time $5.1\times10^3$
sec after the trigger(Beardmore et al. 2011b).

Obviously, After the break at $t\sim8000$ seconds, the optical decay
should be the forward shock emission of the afterglow, at the same
decay slope with that of X-ray afterglow.

In this work we focus on the physical origin of the optical
afterglow emission and pay special attention to the magnetization of
the emitting region. The possible high linear polarization of the
second flare will be discussed in some detail.

\section{The physical origin of the optical emission of GRB110205A}
The optical lightcurve of GRB 110205A is distinguished by two
flares. To interpret the data we divide the observed R-band
lightcurve into three segments: phase-I is the decline phase of the
first flare,  phase II is the the second flare up to $t\sim 8000$ s
(i.e., the break time in the decline) and  phase III is the late
afterglow (see Fig.1).

In this section, we focus on two possible scenarios. In the first scenario,
phase-I is the reverse shock emission of the main outflow, and phase-II is
the optical flare powered by the prolonged activity of the central
engine. In the second scenario, phase-I is the low energy tail of the
prompt soft $\gamma$-ray emission, and phase-II is the reverse shock
emission of the initial outflow. In both scenarios, phase-III
is the regular forward shock afterglow emission.

\begin{figure}[Fig.1]
   \centering
   \includegraphics[width=9.0cm, angle=0]{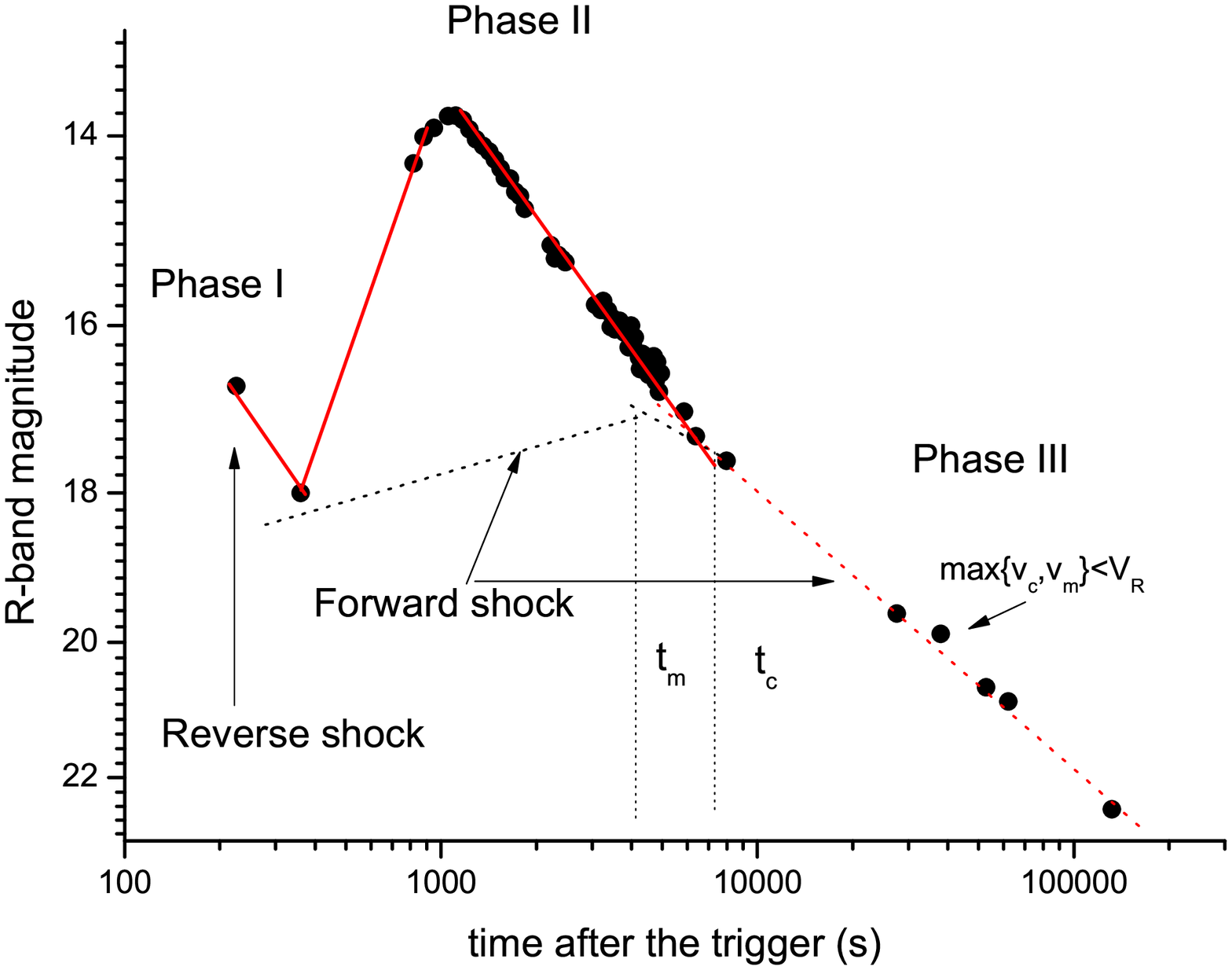}
   \begin{minipage}[]{85mm}
   \caption{The interpretation of the optical emission of GRB 110205A: the first scenario.
  The data points are taken from Klotz et al. (2011a, 2011b), Andreev et al. (2011) and
   Kelemen et al. (2011).} \end{minipage}
   \label{Fig1}
   \end{figure}

\subsection{The first scenario}

The early optical emission may be powered by the reverse shock. The
time when the reverse shock crosses the GRB shell can be estimated
as $\rm t_{\times}{\sim max}[T_{90}, t_{\rm dec} ]$, where $\rm
t_{\rm dec}$ is the deceleration time of the fireball and is given
by $\rm t_{\rm dec}=[(3E_{\rm k}/4\pi\gamma_{0}^2nm_{\rm
p}c^2)^{1/3}/2\gamma_{0}^2c](1+z)$, where $\rm \gamma_{0}$ is the
initial Lorentz factor of the GRB outflow, $n$ is the number density
of the circum-burst medium, $c$ is the speed of light, $m_{\rm p}$
is the rest mass of protons and $E_{\rm k}$ is the
isotropic-equivalent kinetic energy of the GRB outflow, here we
assume $E_{\rm k}=5E_{\rm iso}$, corresponding to a moderate GRB
efficiency $\eta_{\gamma}\equiv E_{\rm iso}/(E_{\rm iso}+E_{\rm
k})\approx 0.17$. The thick shell case corresponds to $T_{90}\sim
t_{\times}$ while the thin shell case corresponds to $T_{90}<
t_{\times}$. For the current burst, the reverse shock emission peaks
at $\rm t_{\rm p,r}=225$ sec, where the superscript 'r'  represents
the parameter of the reverse shock, hence we have $t_{\times}\sim
t_{\rm p,r}\sim T_{90}\sim t_{\rm dec}$ and
$\gamma_{0}\sim[\frac{24E_{\rm k}(1+z)^3}{\pi n m_{\rm p}c^5 t_{\rm
p,r}^3}]^{1/8}=4.3\times10^2E_{\rm k,54.3}^{1/8}n_{0}^{-1/8}t_{\rm
p,r,2.4}^{-3/8} (\frac{1+z}{3.2})^{3/8}$.

When reverse shock crosses the thick shell its emission is governed
by Sari \& Piran (1999) and Kobayashi (2000)
\begin{equation}
\nu_{\rm m}^{\rm r}(t_{\times})\approx5.0\times 10^{13}
{\epsilon_{\rm e,-1}^{\rm r}}^{2}
{\epsilon_{\rm B,-2}^{\rm r}}^{1/2}n_{0}^{1/2}
\gamma_{0,2.6}^{2}(\frac{1+z}{3.2})^{-1}~{\rm Hz},
\end{equation}

\begin{equation}
\nu_{\rm c}^{\rm r}(t_{\times})\approx8.5\times10^{14}
{\epsilon_{\rm B,-2}^{\rm r}}^{-3/2}
E_{\rm k,54.3}^{-1/2}n_{0}^{-1}(\frac{1+z}{3.2})^{-1/2}t_{\times,2.4}^{-1/2}
~{\rm Hz},
\end{equation}

\begin{equation}
F_{\nu, \rm max}^{\rm r}(t_{\times})\approx 5.1\times10^{2}
{\epsilon_{\rm B,-2}^{\rm r}}^{1/2}E_{\rm k,54.3}^{5/4}n_{0}^{1/4}
\gamma_{0,2.6}^{-1}t_{\times,2.4}^{-3/4}(\frac{1+z}{3.2})
D_{\rm L,28.7}^{-2} ~{\rm mJy},
\end{equation}
where $\epsilon_{\rm B}$ and $\epsilon_{\rm e}$ are the fractions of
the shock energy given to the magnetic field and electrons at the
shock front respectively, and $D_{\rm L}$ is the luminosity
distance. The synchrotron emission for $\nu_{\rm m}^{\rm r}<\nu_{\rm
R}<\nu_{\rm c}^{\rm r}$ is estimated by $F_{\nu_{\rm R}}^{\rm
r}=F_{\nu, \rm max}^{r} (\nu_{\rm R}/\nu_{\rm m}^{\rm r})^{\rm
-(p-1)/2}\approx 50 {\epsilon_{\rm B,-2}^{\rm r}}{\epsilon_{\rm
e,-1}^{\rm r}}^{2}E_{\rm k,
54.3}^{5/4}n_{0}^{3/4}\gamma_{0,2.6}t_{\times,2.4}^{-3/4} D_{\rm
L,28.7}^{-2}$ mJy,  where $p$ is the power-law index of electrons
distribution (i.e., $N_{\rm \gamma}\propto \gamma^{-p}$). For
$t>t_{\times}$, the reverse shock emission drops with time as
$F_{\nu}^{\rm r} \propto t^{\rm -(73p+21)/96}$ (Kobayashi 2000, Fan
et al. 2002). Taking $p=3$, we can get $\rm F_{\nu}^{r}\propto
t^{-2.5}$, matching the decline of the first flare. The reverse
shock optical emission flux can be consistent
 with the observation for proper parameters, such as $\epsilon_{\rm e}^{\rm r}\sim 0.1$, $\epsilon_{\rm
B,-2}^{\rm r}\sim 0.1$ and $n_{0}\sim 10^{-2}$.

{\it It is however important to check whether the late ($t>0.1$ day)
afterglow emission can be interpreted as the forward shock
emission.} For $t>0.1$ day, the X-ray and optical afterglow data
decline with the time as $t^{-1.6}$, suggesting that the R-band is
above both the typical synchrotron radiation frequency ($\nu_{\rm
m}$) and the cooling frequency ($\nu_{\rm c}$) of the forward shock
electrons. Therefore both $\nu_c$ and $\nu_m$ should cross the
observer's frequency ($\nu_{\rm R}$) before $t<0.1$ day. These two
times are denoted as $t_{\rm c}$ and $t_{\rm m}$, respectively.

For the ISM-like medium\footnote{ In the case of the free
stellar-wind medium, the decline of the forward shock emission after
the peak of the reverse shock can not be shallower than $t^{-1/4}$
(e.g., see Tab.1 of Xue et al. 2009). On the other hand the forward
shock R-band emission should be dimmer than $18^{\rm th}$ mag at
$t=350$ s and then be the same as that detected at $t\geq 0.1$ day.
With Fig.1 it is straightforward to show that such requests can not
be satisfied. So we do not discuss the wind medium model here.} and
$t>t_{\rm dec}$, we have $\rm F_{\nu,max}\propto t^{0}$, $\rm
\nu_{m}\propto t^{-3/2}$ and $\rm \nu_{c}\propto t^{-1/2}$ (Sari \&
Piran 1999). In the fast cooling case, the forward shock emission
can be estimated as
\begin{equation}
\rm F_\nu=\cases{ ( \nu / \nu_c )^{1/3} F_{\nu,max}\propto t^{1/6}, &
$\rm \nu_{m}>\nu_c>\nu$, \cr ( \nu / \nu_c )^{-1/2} F_{\nu,max}\propto t^{-1/4}, &
$\nu_m>\nu>\nu_c$, \cr ( \nu_m / \nu_c )^{-1/2} ( \nu / \nu_m)^{-p/2}
F_{\nu,max}\propto t^{-3/2}, & $\nu>\nu_m>\nu_{c}$, \cr }
\label{eq:2.1-1}
\end{equation}
where $p\sim 2.6$ has been adopted \footnote{Such a value is roughly
consistent with that needed to account for the late afterglow
decline ($F \propto t^{-1.6}$) and that required to reproduce the
late time X-ray spectrum ($F_\nu \propto \nu^{-1.15\pm 0.08}$).}.
Evidently, the forward shock R-band emission should be dimmer than
$18^{\rm th}$ mag at $t=350$ s and then be the same as that detected
at $t>0.1$ day. It is straightforward to show that it is not
possible to find proper $t_{\rm c}$ and $t_{\rm m}$ with the scaling
laws of eq.(\ref{eq:2.1-1}). Hence this possibility has been ruled
out.

In the slow cooling case,
\begin{equation}
\rm F_\nu=\cases{(\nu/\nu_m)^{1/3} F_{\nu,max}\propto t^{1/2},
            & $\nu_{c}>\nu_m>\nu$, \cr
(\nu/\nu_m)^{-(p-1)/2} F_{\nu,max}\propto t^{-6/5},& $\nu_c>\nu>\nu_m$, \cr
\left( \nu_c/\nu_m \right)^{-(p-1)/2}
\left( \nu/\nu_c \right)^{-p/2} F_{\nu,max}\propto t^{-3/2},
            & $\nu>\nu_c>\nu_{m}$. \cr}
\label{eq:2.1-2}
\end{equation}
In this case one can find proper $t_{\rm c}\sim 4\times 10^{3}$ s
and $t_{\rm m}\sim 8\times 10^{3}$ s, as illustrated in Fig.1. Hence
it deserves a further investigation. For $t> t_{\rm dec}$, the
dynamics of the forward shock can be well approximated by the
Blandford-McKee similar solution (Blandford \& McKee 1976), which
emission can be estimated by
\begin{equation}
F_{\nu,{\rm max}}=23~{\rm mJy}~({1+z\over 3.2}) D_{\rm L,28.7}^{-2}
\epsilon_{\rm B,-2}^{1/2}E_{\rm k,54.3}n_{0}^{1/2},
\end{equation}
\begin{equation}
\nu_{\rm m} =9.0\times 10^{15}~{\rm Hz}~E_{\rm k,54.3}^{1/2}\epsilon_{\rm
B,-2}^{1/2}\epsilon_{e,-1}^2 ({C_{\rm p}\over1.6})^2 ({1+z \over 3.2})^{1/2}
t_{3}^{-3/2},
\end{equation}
\begin{equation}
\nu_{\rm c} = 2.3\times 10^{15}~{\rm Hz}~E_{\rm k,
54.3}^{-1/2}\epsilon_{\rm B,-2}^{-3/2}n_{0}^{-1}
 ({1+z \over 3.2})^{-1/2}t_{3}^{-1/2},
 \end{equation}
where $C_p \equiv 13(p-2)/[3(p-1)]$. As shown in Fig.1, the data
suggest that $F_{\nu,{\rm max}} \sim 0.4~{\rm mJy}$ i.e.,
$\epsilon_{\rm B,-2}^{1/2}E_{\rm k,54.3}n_{0}^{1/2} \sim 1/60$. The
fact that $\nu_{\rm m}$ and $\nu_{\rm c}$ crosses $\nu_{\rm R}$ at
$t_{\rm m}$ and $t_{\rm c}$ gives that $E_{\rm
k,54.3}^{1/2}\epsilon_{\rm B,-2}^{1/2}\epsilon_{e,-1}^2 \sim 0.5$
and $E_{\rm k, 54.3}^{-1/2}\epsilon_{\rm B,-2}^{-3/2}n_{0}^{-1}\sim
0.6$. To reproduce the data, one needs $\epsilon_{\rm
B,-2}\sim4\times 10^{7}E_{\rm k,54.3}^{3}$, which is hard to satisfy
for reasonable GRB efficiency. Let's check whether the inverse
compton effect can change the result or not. With such an effect,
$\nu_{\rm c}\propto (1+Y)^{-2}$, where $Y$ is the compton parameter
and can be estimated by $Y\sim(-1+\sqrt{1+4\eta \epsilon_{\rm
e}/\epsilon_{\rm B}})/2$, where $\eta\sim \rm min\{1,[\nu_{\rm
m}/(1+Y)^{2}\nu_{\rm c}]^{(p-2)/2}\}$. For $t\sim t_{\rm c}\sim
t_{\rm m}$, we have $\nu_{\rm c} \sim \nu_{\rm m}$, $\eta \sim
(1+Y)^{2-p}$ and $Y \sim(\epsilon_{\rm e}/\epsilon_{\rm B})^{0.4}$
if $\eta \epsilon_{\rm e}/\epsilon_{\rm B} \gg
 1$. So the cooling frequency is estimated by $\nu_{\rm c} = 3.6\times
10^{14}~{\rm Hz}~E_{\rm k,54.3}^{-1/2}\epsilon_{\rm
e,-1}^{-0.8}\epsilon_{\rm B,-2}^{-0.7}n_{0}^{-1} ({1+z \over
3.2})^{-1/2}t_{3}^{-1/2}$ (for $t_{\rm c} \sim 8\times 10^{3}$ s we
have $E_{\rm k, 54.3}^{-1/2}\epsilon_{\rm e,-1}^{-0.8}\epsilon_{\rm
B,-2}^{-0.7}n_{0}^{-1}\sim 4$). In this case we need $\epsilon_{\rm
B}\sim 6\times 10^{-9}E_{\rm k,54.3}^{-3.4}$ to reproduce the data,
which seems too small to be reasonable. A reasonable $\epsilon_{\rm
B} \sim 10^{-3}-10^{-2}$ is obtainable if we take $E_{\rm k}\sim
E_{\rm iso}/20$, i.e., the GRB efficiency is as high as $\eta_\gamma \sim 0.95$. It
is unclear how such a high GRB efficiency can be achieved. Indeed
very high $\eta_\gamma$ have been reported for some Swift GRBs in
the literature (e.g., Zhang et al. 2007; Cenko et al. 2010). One,
however, should take these preliminary results with caution. For
example, in the energy injection model for the X-ray shallow decline
phase, a very high $\eta_\gamma$ can be obtained if one estimates
$E_{\rm k}$ with the shallow decline data. However, as pointed out
by Fan \& Piran (2006) for the first time, in many cases the energy
injection model can not interpret the simultaneous optical data
self-consistently. Therefore the very high GRB efficiency found in
the specific energy injection model is questionable. For some
bursts, for instance GRB 080319B, the prompt gamma-ray emission
plausibly came from a narrow energetic core while the late time
afterglow emission was powered by a wider but much less powerful
ejecta (e.g., Racusin et al. 2008). The derived GRB efficiency would
be unreasonably high if one estimates $E_{\rm k}$ with the late
afterglow data rather than the early afterglow data. Considering
these uncertainties, we do not examine the ultra-high $\eta_\gamma$
model further.

Since neither the slow nor the fast cooling case can provide a
self-consistent interpretation of the forward shock emission, we
conclude that the first scenario is not favored.

\subsection{The second scenario}

BAT observation showed that the prompt soft gamma-ray emission with
multiple peaks lasted to $>300$ seconds after the trigger. That
means when the first optical flare peaked at about 225 seconds, the
prompt gamma-ray emission was still ongoing. So it is possible that
phase-I is the low energy tail of prompt emission. Indeed bright
prompt optical emission has been well detected in some GRBs, such as
GRB 041219A (Blake et al. 2005) and GRB 080319B (Racusin et al.
2008). In this case phase-II may be powered by the reverse shock of
the initial prompt GRB. If correct, the fireball is thin since the
reverse shock crossed the outflow at a time $t_\times \sim 1140~{\rm
s} \gg T_{\rm 90}$. The initial Lorentz factor of the GRB outflow
can be estimated
\begin{equation}
\gamma_{0}\sim[\frac{24E_{\rm k}(1+z)^3}{\pi n m_{\rm p}c^5 t_{\rm
\times}^3}]^{1/8}=2.8\times10^2E_{\rm
k,54.3}^{1/8}n_{0}^{-1/8}t_{\times,3.1}^{-3/8}
(\frac{1+z}{3.2})^{3/8}.
\end{equation}

In the thin shell case, for $\nu_{\rm m}^{\rm r}<\nu_{\rm
R}<\nu_{\rm c}^{\rm r}$ and $t<t_{\times}$, $F_{\nu_{\rm R}}^{\rm
r}\propto t^{2p-1}\propto t^{4.2}$ (e.g., Fan et al. 2002),
consistent with the rise behavior of phase-II $F_{\nu_{\rm
R}}\propto t^{4}$. {\it The decline of phase-II is expected to be
$F_{\nu_{\rm R}}\propto t^{-(27p+5)/35}$ if it is dominated by the
synchrotron radiation of the cooling reverse shock electrons or
alternatively $F_{\nu_{\rm R}}\propto t^{-[2+(p-1)/2]}$ if dominated
by the high latitude emission of the reverse shock, in agreement
with the observation for $p\sim 2.6$ (we'll return to this point
later).} Therefore both the increase and the decline of phase-II can
be reasonably interpreted within the reverse shock model.

Following Fan et al. (2002), the peak optical emission of reverse
shock can be estimated as

\begin{equation} F_{\nu_{\rm R}}^{\rm
r}\approx 3~{\rm mJy}~[8(p-2)/3(p-1)]^{p-1}{\epsilon_{\rm e,-1}^{\rm
r}}^{p-1}{\epsilon_{\rm B,-3}^{\rm
r}}^{(p+1)/4}n_0^{(p+1)/4}(\gamma_0/280)^{p+1},
\end{equation}
where the total number of the reverse-shock-accelerated electrons
$\sim E_{\rm k}/\gamma_0 m_{\rm p}c^{2}$ and the strength of reverse
shock $\Gamma_{\rm rs} \sim 1.3$ have been adopted.  For proper
parameters, for example $\epsilon_{\rm B}^{\rm r}\sim \epsilon_{\rm
e}^{\rm r}\sim 0.1$ and $n_0\sim 1$, the reverse shock emission is
so bright that can account for the observation data.

As in the first scenario, it is important to check whether the
forward shock emission can account for the late afterglow data. In
principle there could be three regimes.

Case I: $\nu_{\rm R}>\rm max\{\nu_{\rm m}(t_{\times}),\nu_{\rm
c}(t_{\times})\}$, for which the temporal behaviors of the forward
shock optical emission are given by (e.g., Xue et al. 2009)
\begin{equation}
F_{\nu_{\rm R}} \propto \cases{ t^{2}, &
for~~~$t<t_\times$, \cr t^{-1.5}, &
for~~~$t>t_\times$.}
\label{eq:2.2-1}
\end{equation}

Case II: $\nu_{\rm m}(t_{\times})<\nu_{\rm R}<\nu_{\rm
c}(t_{\times})$. In this case $t_{\rm c}=2.1\times10^{4}\rm
sec~E_{\rm k, 54.3}^{-1}\epsilon_{\rm B,-2}^{-3}n_{0}^{-2}({1+z
\over 3.2})^{-1}$, at which $\nu_{\rm c}$ crosses the observer's
frequency, should be introduced. The temporal behaviors of the
forward shock optical emission are given by (e.g., Xue et al. 2009)
\begin{equation}
F_{\nu_{\rm R}} \propto \cases{
t^{3}, & for~~~$t<t_\times$, \cr
t^{-1.2}, & for~~~$t_\times<t<t_{\rm c}$, \cr
t^{-1.5}, & for~~~$t>t_{\rm c}$.}
\label{eq:2.2-2}
\end{equation}

Case III: $\nu_{\rm c}(t_{\times})<\nu_{\rm R}<\nu_{\rm
m}(t_{\times})$. In this case we introduce $t_{\rm
m}=6.8\times10^{3}~{\rm  sec}~E_{\rm k, 54.3}^{1/3}\epsilon_{\rm
B,-2}^{1/3}\epsilon_{\rm e,-1}^{4/3} ({1+z \over 3.2})^{1/3}$ when
$\nu_{\rm m}$ crosses the observer's frequency. The temporal
behaviors of the forward shock optical emission are given by (e.g.,
Xue et al. 2009)
\begin{equation}
F_{\nu_{\rm R}} \propto \cases{
t^{2}, & for~~~$t<t_\times$, \cr
t^{-1/4}, & for~~~$t_\times<t<t_{\rm m}$, \cr
t^{-1.5}, & for~~~$t>t_{\rm m}$.}
\label{eq:2.2-3}
\end{equation}

Since the forward shock optical emission should be dimmer than $\sim
18^{\rm th}$ mag at $t\sim 350$ s and should be consistent with the
afterglow data for $t>0.1$ day. With the temporal behaviors
suggested in Case-I, the predicted optical emission at $t\sim 350$ s
is much brighter than $\sim 18^{\rm th}$ mag, rending this
possibility unlikely.

Usually the cooling frequency of the forward shock emission is
comparable to or larger than than that of the reverse shock. In the
modeling of the temporal behaviors of phase-II, the cooling
frequency of the reverse shock is required to be above $\nu_{\rm
R}$. Therefore Case-III is disfavored and just Case-II remains. As
shown in Fig.2, Case-II works as long as $t_{\rm c}\leq 8\times
10^{3}$ s. After the reverse shock crossed the ejecta, the cooling
frequency of the reverse shock electrons drops with time as
$t^{-3/2}$ (Sari \& Piran 1999). Therefore $\nu_{\rm
c}^{r}(t_\times)\geq (t_{\rm c}/t_\times)^{3/2}\nu_{\rm R}$ is
needed if the quickly decaying optical emission is the synchrotron
radiation of the cooling reverse shock electrons (hereafter {\it the
synchrotron radiation case}). Alternatively $\nu_{\rm
c}^{r}(t_\times)\geq (t_{\rm c}/t_\times)\nu_{\rm R}$ is needed if
the quickly decaying optical emission is dominated by the high
latitude emission of the reverse shock (hereafter {\it the high
latitude emission case}). However, we have $\nu_{\rm
c}(t_\times)\approx (t_{\rm c}/t_\times)^{1/2}\nu_{\rm R}$.
Therefore we need $\nu_{\rm c}^{r}(t_\times) \sim (t_{\rm
c}/t_\times) \nu_{\rm c}(t_\times)$ or $\nu_{\rm c}^{r}(t_\times)
\sim (t_{\rm c}/t_\times)^{1/2}\nu_{\rm c}(t_\times)$. Following Jin
\& Fan (2007) we have $\nu_{\rm c}^{r}(t_\times) \sim \Re_{\rm
B}^{-3}(\frac{1+Y}{1+Y^{\rm r}})^2\nu_{\rm c}(t_\times)$, where
$Y^{\rm r}>0$ is the Compton parameter of the reverse shock. Note
that $\nu_{\rm m}^{\rm r}\sim 3\times 10^{11}~{\rm Hz}~
{\epsilon_{\rm e,-1}^{\rm r}}^{2}{\epsilon_{\rm B,-2}^{\rm
r}}^{1/2}n_0^{1/2}\ll \nu_{\rm c}^{\rm r}$ unless ${\epsilon_{\rm
e,-1}^{\rm r}}\equiv {\Re_{\rm e}\epsilon_{\rm e,-1}}\sim 10$ (Fan
et al. 2002), therefore $Y^{\rm r}\leq 1$ for $(\nu_{\rm m}^{\rm
r}/\nu_{\rm c}^{\rm r})^{(2-p)/2}(\epsilon_{\rm e}^{\rm
r}/\epsilon_{\rm B}^{\rm r})\leq 1$. Hence we have $\nu_{\rm
c}^{r}(t_{\times}) \sim \Re_{\rm B}^{-3}(1+Y)^2\nu_{\rm
c}(t_{\times})$, which further gives that ${\Re_{\rm B}}\sim
0.5(1+Y)^{2/3}$ (synchrotron radiation case) or ${\Re_{\rm B}}\sim
0.7(1+Y)^{2/3}$ (high latitude emission case), implying the outflow
could be (weakly) magnetized for $Y\sim (\epsilon_{\rm
e}/\epsilon_{\rm B})^{0.4} \geq {\rm a ~few}$.

A magnetized outflow is also consistent with the flux analysis. In
Case-II, the ratio between the reverse shock optical emission and
forward shock optical emission at the crossing time of the reverse
shock can be estimated by (Jin \& Fan 2007; Fan et al. 2002; Zhang
et al. 2003)
\begin{equation}
\frac{\rm F_{\nu_{\rm R}}^{r}(t_{\times})}{\rm F_{\nu_{\rm R}}(t_{\times})}\approx
0.08\Re_{\rm e}^{p-1}\Re_{\rm B}^{\frac{p+1}{2}}=0.08\Re_{\rm e}^{1.6}\Re_{\rm B}^{1.8}.
\end{equation}

The current observation suggests that $F_{\nu_{\rm
R}}^{r}(t_{\times})/F_{\nu_{\rm R}}(t_{\times})\geq 4$, we then have
$\Re_{\rm e}^{1.6}\Re_{\rm B}^{1.8}\geq 50$, i.e.,
\begin{equation}
\Re_{\rm B}\geq 8.8\Re_{\rm e}^{-0.9}.
\end{equation}
For $\Re_{\rm e}\leq 10$ we have $\Re_{\rm B}\geq 1$
unless phase-II was dominated by the emission of prolonged activity
of the central engine (e.g., Gao 2009).

\begin{figure}[Fig.2]
   \centering
   \includegraphics[width=9.0cm, angle=0]{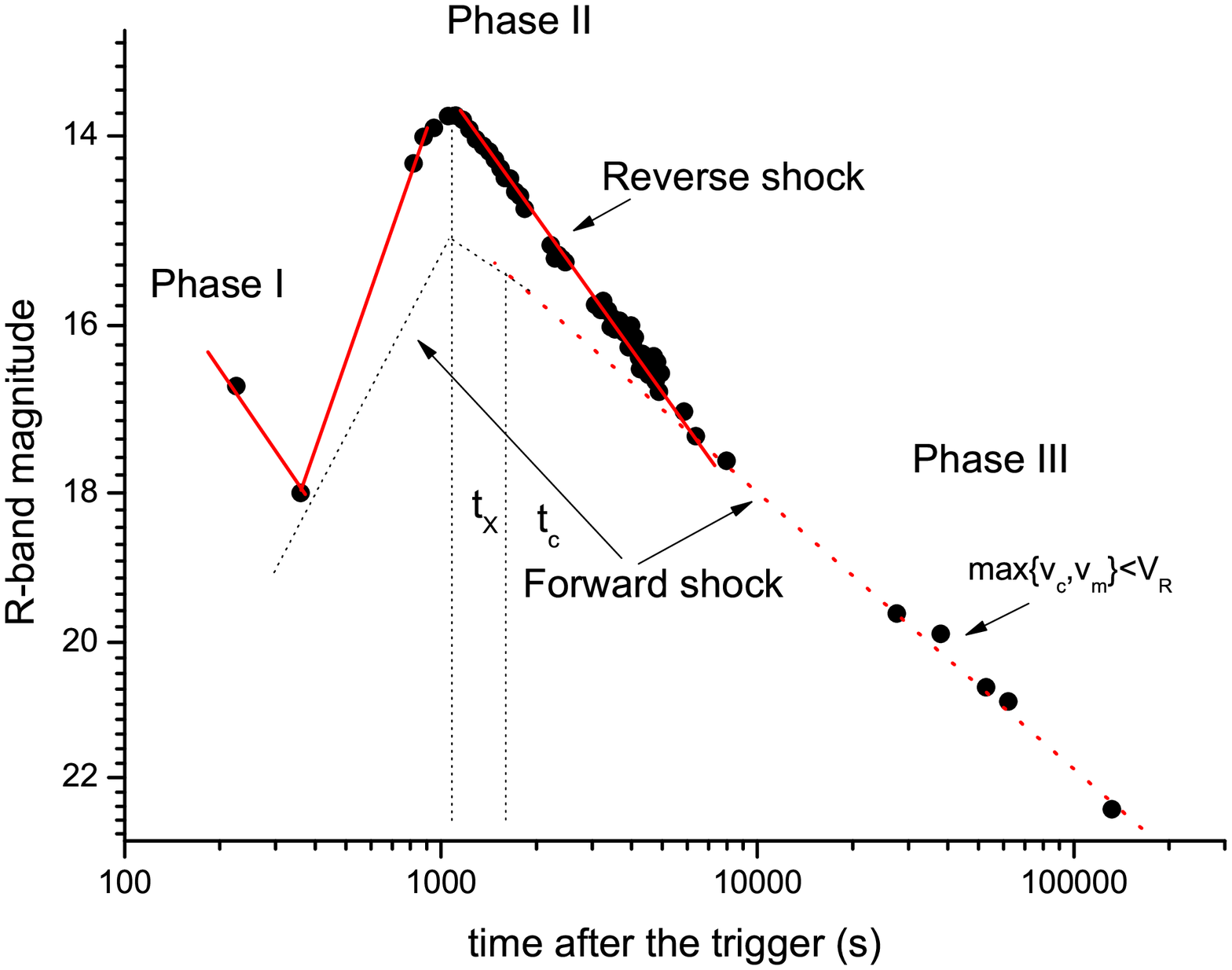}
   \begin{minipage}[]{85mm}
   \caption{The interpretation of the optical
   emission of GRB 110205A: the second scenario.}
   \end{minipage}
   \label{Fig2}
   \end{figure}

\section{The expected polarization property of the optical flares}
The polarimetry of prompt emission or the reverse shock emission is
very important for diagnosing the composition of the GRB outflow,
since the late afterglow, taking place hours after the trigger, is
powered by the external forward shock, so that essentially all the
initial information about the ejecta has lost. If the GRB ejecta is
initially magnetized, the prompt $\gamma-$ray/X-ray/UV/optical
emission and the reverse shock emission should be linearly polarized
(Lyutikov et al. 2003; Granot 2003; Fan et al. 2004). In a few
events, the prompt $\gamma-$ray polarimetry are available but the
results are quite uncertain. Even for the very bright GRB 041219A, a
systematic effect that could mimic the weak polarization signal
could not be excluded \citep{McGl07}. So far, the most reliable
polarimetry is that in UV/optical band (e.g., Covino et al. 1999).
The optical polarimetry of the prompt emission or the reverse shock
emission demands a quick response of the telescope to the GRB alert
and is thus very challenging. Nevertheless, significant progress
have been made since 2006. Using a ring polarimeter on the robotic
Liverpool Telescope, Mundell et al. (2007) got an upper limit $\sim
8\%$ of the optical polarization of the afterglow of GRB 060418 at
203 sec after the trigger. A breakthrough was made in GRB 090102,
for which Steele et al. (2009) found out that the reverse shock
emission was linearly polarized with a degree $\sim 10.1\%$. The
forward-reverse shock modeling of the afterglow emission suggests a
mildly magnetized reverse shock region (Fan 2009; Mimica et al. 2010
). These two findings are consistent with each other and provide the
most compelling evidence so far for the magnetized outflow model for
GRBs.

For the current event, we have shown that phase-I may be the low
energy tail of the prompt emission and phase-II is likely the
reverse shock emission though other possibility---the prolonged
activity of the central engine can not be ruled out. In the reverse
shock model we find out that the initial outflow could be weakly
magnetized. Assuming that in the emitting region the strength ratio
of the ordered magnetic field and the random magnetic field is $b$,
the degree of net polarization can be roughly estimated as $\Pi_{\rm
net}\approx 0.6 b^2/(1+b^2)$ (e.g., Fan et al. 2004). Therefore a
moderate or high linear-polarization-degree of the weakly magnetized
reverse shock emission is possible. In reality the configuration of
the outflowing magnetic field may be very complicated and the
direction of the lines may be not well ordered, for which the net
polarization degree should be lowered. In section 2.2 we have
shown that the decline of the second optical flare may be
dominated by the high latitude emission of the reverse shock. In
such a case, {\it an interesting polarization degree evolution presents
if at late times (i.e., $t>t_\times$) the edge of the ejecta is
visible} (see Fig.2 of Fan et al. 2008 for illustration).
Alternatively if the decline of the second optical flare is shaped
by the synchrotron radiation of the cooling reverse shock electrons,
the observed linear polarization should be $
\Pi=(\Pi^{r}F_{\nu_{_{\rm R}}}^r+\Pi F_{\nu_{_{\rm
R}}})/(F_{\nu_{_{\rm R}}}^r+F_{\nu_{_{\rm R}}})$, where $\Pi \sim
1.4\%$ ($\Pi^{r}$) is the linear polarization of the forward
(reverse) shock emission (Gorosabel et al. 2011), and $F_{\nu_{_{\rm
R}}}$ ($F_{\nu_{_{\rm R}}}^{\rm r}$) is the R-band flux of the
forward (reverse) shock emission. {\it One thus expects a steady decline of the
polarization degree with the time.}

The Liverpool Telescope began automatically observing Swift GRB
110205A with the RINGO2 polarimeter (Mundell et al. 2011). RINGO2 is
quite sensitive and can measure the linear polarization degree down
to 0.9$\%$ for the source as bright as $R=17$ mag with the exposure
time of 100 seconds. Therefore the linear polarization of first
optical flare might be detectable if it is significantly polarized.
For the second optical flare, the emission is so bright that the
time-resolved polarization property can be achieved. The
observations of Liverpool Telescope on GRB 110205A have stopped due
to bad weather and the polarization data have not been published
yet. Nevertheless the current optical data have shed some light on
the nature of the GRB outflow and have suggested very interesting
polarization properties.

\section{Conclusion and Discussion}
The Early optical emission of GRB 110205A is characterized by two
flares. Two possible scenarios have been examined in this work. In
the first scenario, the first optical flare is the reverse shock
emission of the main outflow, the second flare is the emission
powered by the prolonged activity of the central engine, and the
late time ($t>0.1$ day) optical emission is the regular forward
shock emission. However we find that a self-consistent
interpretation of the forward shock emission is hard to achieve
unless the GRB efficiency is $\sim 0.95$. Therefore this scenario is
disfavored. In the second scenario, the first optical flare is the
low energy prompt emission while the second flare is the reverse
shock emission of the main outflow. The late optical emission is
still attributed to the regular forward shock. Within this scenario
a self-consistent interpretation is achieved (please note that the
different Inverse Compton parameters for the reverse and forward
socks play an important role in the modeling). There are two
interesting findings. One is that the reverse shock region may be
weakly magnetized and a moderate linear polarization degree is
likely. The other is that the decline of the second optical flare
may be dominated by the high latitude emission. Similar to the X-ray
case, {\it one may expect dramatic polarization evolution} (Fan et
al. 2008) accompanying the optical flare decline rather than the
steady decrease of the polarization degree. In reality the
configuration of the out-flowing magnetic field may be very
complicated and the direction of the lines may be not well ordered,
for which the net polarization degree should be lowered and the
polarization evolution may be very complicated.

The Liverpool Telescope began automatically observing Swift GRB
110205A with the RINGO2 polarimeter (Mundell et al. 2011). RINGO2 is
very sensitive. The linear polarization of first optical flare may
be detectable if it is significantly polarized. For the second
optical flare, the emission is so bright that the time-resolved
polarization properties can be achieved. Therefore our prediction,
in particular strong polarization evolution in the decline phase of
the second optical flare, can be directly tested.

\normalem
\begin{acknowledgements}
We thank the anonymous referee for helpful comments and Dr. Yi-Zhong
Fan for help in improving the presentation. This work was supported
by the National Natural Science Foundation of China under the grant
11073057.
\end{acknowledgements}

\clearpage

\end{document}